# Stochastic differential equation for modelling health related quality of life

Ralph Brinks

## Abstract

In this work we propose a stochastic differential equation (SDE) for modelling health related quality of life (HRQoL) over a lifespan. HRQoL is assumed to be bounded between 0 and 1, equivalent to death and perfect health, respectively. Drift and diffusion parameters of the SDE are chosen to mimic decreasing HRQoL over life and ensuring epidemiological meaningfulness. The Euler-Maruyama method is used to simulate trajectories of individuals in a population of n = 1000 people. Age of death of an individual is simulated as a stopping time with Weibull distribution conditioning the current value of HRQoL as time-varying covariate. The life expectancy and health adjusted life years are compared to the corresponding values for German women.

## Introduction

In the classical theory of compression of morbidity, periods of life with and without disease are compared with life expectancy [Fri83]. While compression of morbidity is well expressed in terms of the illness-death model (IDM) [Bri18], health related quality of life (HRQoL) does not play a role in the IDM. Missing HRQoL may lead to a paradoxical situation: for some diseases, there is an expansion of morbidity (which is seen as disadvantageous) but the overall burden of the disease is decreasing (which is seen as an advantage). The reason for this paradox might be seen in medical progress such that diseases with formerly burdensome treatments can be dealt with less adverse effects. Moreover, in using the IDM we frequently only deal with one medical condition: the condition is either present or not. The compression theory of Fries introduced the term morbidity and the question arises, what morbidity is. It is doubtful whether an individual is either morbid or not. A more realistic and modern picture seems to be that health is a continuum [Key16]. Indeed, HRQoL takes values from a continuum. Taking both points together, paradox observations about some diseases and the idea of health being a continuum, it seems plausible that for a comprehensive description of the health situation in a population, the IDM is not optimal and HRQoL needs to be taken into account.

In this article we consider the idea that HRQoL of an individual can be seen as a stochastic process $X := (X_t, t \in I)$ parameterized by age $t$ over the lifespan $I$. Reasons for modelling HRQoL as a stochastic process lie in the highly dynamic time-dependency of the patients' individual physical, mental, functional and social abilities - and possibly most important - the individuals' ratings and feelings about these abilities. For example, a mediocre health situation can be perceived *subjectively* very differently from one day to the other, although *objectively* little has changed. Since we believe that HRQoL should take into account the subjective ratings, HRQoL can vary substantially from one day to another. Thus, we think that HRQoL can be modelled stochastically. The idea of the stochastic nature of HRQoL is not new [Sch91, Pal96], however, to our knowledge HRQoL has never been modelled as a stochastic process, which is the solution of an Ito SDE.

## Methods

We set up an Ito stochastic differential equation (SDE) for modelling HRQoL over the lifespan of an individual. The time variable $t \in I := [0, \omega] \subset [0, \infty)$ of the underlying stochastic process $X := (X_t, t \in I)$ represents the age of the individual. $X_t, t \in I,$ is the HRQoL at age $t$, which is assumed to be bounded between 0 and 1, $0 \leq X_t \leq 1, t \in I$. Values $X_t = 0$ and $X_t = 1$ are equivalent to death and perfect health, respectively. The Ito SDE has the form

$$dX_t = b(t, X_t)\,dt + \sigma(t, X_t)\,dB_t, \qquad (1)$$

with drift $b$ and diffusion $\sigma$. Drift $b$ and diffusion $\sigma$ are chosen such that (on average) $X_t$ is decreasing and meaningful, i.e., $0 \leq X_t \leq 1$ for all $t \in I$. The Euler-Maruyama method [Klo92] is used to simulate trajectories of individuals in a population of $n = 1000$ people with step size $\Delta_t = 0.01$ (years). Age of death of an individual is simulated as a stopping time τ with Weibull distribution conditioning on the current value $X_t$. In this sense, the value $X_t$ of the HRQoL at time $t$ is a time-varying covariate. After the stopping time τ, HRQoL is set to zero, $X_t = 0$ for $t \geq τ$. Let $X^{(τ)}$ denote the stopped process.
In order to describe the population, summary statistics for HRQoL adjusted life years of each subject in the population are calculated by integrating the $n=1000$ paths of the stopped process $X^{(τ)}$. As we use stopping times as upper bounds of these integrals, we note that the constructed process is progressively measurable [Kar98]. We refer to the integral of a path of the stopped process as a health adjusted life year (HALY).

Simulations are performed with the open statistical software R (The R Foundation for Statistical Computing), version 4.1.2, which is distributed under the terms of the GNU General Public License (GNU GPL). The R software is run under Linux Mint 64-bit version 21.1.

# Results

## Choice of parameters

### Drift and diffusion

Choice of drift and diffusion are motivated from the Doering-Cai-Lin family of SDEs [Dom20]. The Doering-Cai-Lin family of stochastic processes has the advantage of being bound from below and above. For Equation (1), we choose as drift $b(t, X_t) = X_t \, \delta(t)$ with $\delta(t)$ as shown in Figure 1 and as diffusion term $\sigma(t, X_t) = 0.05 \, \text{sqrt}\{(1 - X_t) \, X_t\}$. The drift $b$ is the product of the current HRQoL $X_t$ and an age-dependent term $\delta$, which gradually decreases until about age 85 and then shows a strongly decreasing shape for ages above 85. The choice of the diffusion $\sigma$ ensures that $0 \leq X_t \leq 1$ for all $t \in I$.

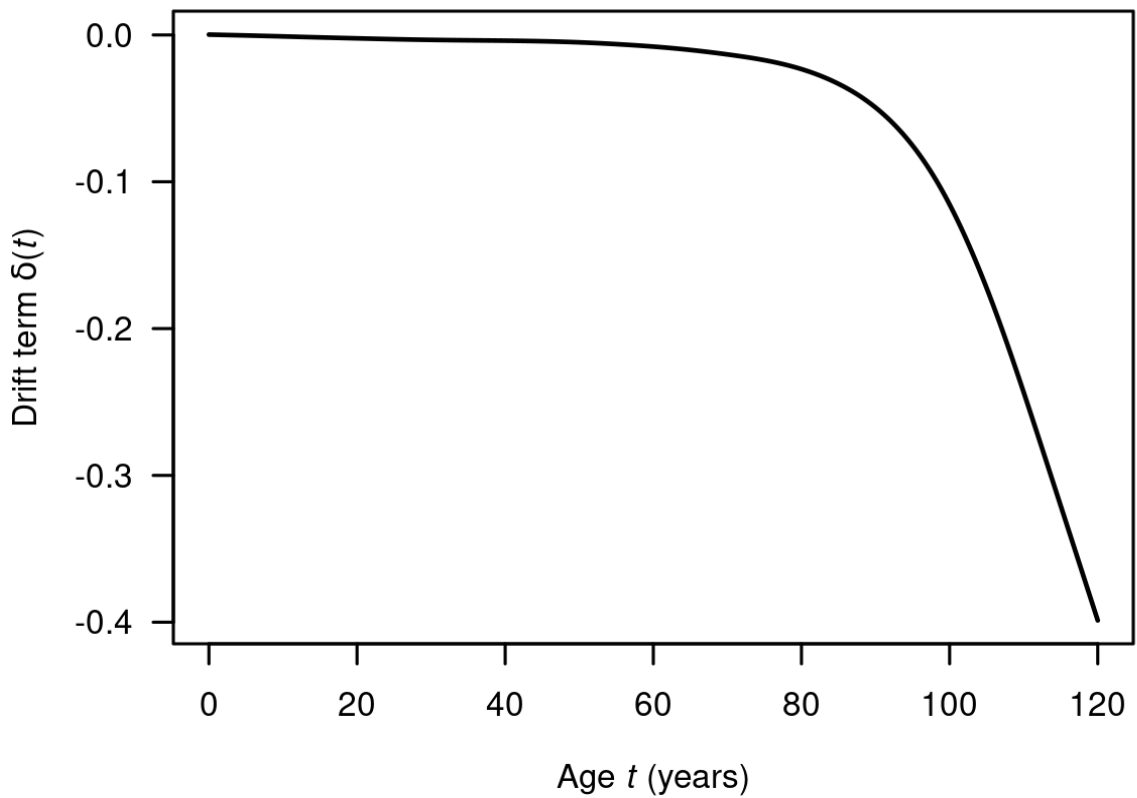

Figure 1: Part of the drift $\delta(t)$ over age $t$.

### Stopping time

Hazard rate of the Weibull distribution for simulating the stopping time is $h_0(t) = \exp(\alpha + \beta \, t)$ with $\alpha = -11.175$ and $\beta = 0.1$, which is motivated by the mortality rate of the general mortality in Germany. As we are conditioning on the current value of $X_t$, we choose the hazard rate

$h(t)$ for an individual to be $h(t) = h_0(t)/\sqrt{X_t}$. Age of death τ is simulated as stopping time by the inverse sampling of the cumulative distribution function $F$ associated with $h(t)$.

## Software

The source code for the findings of this paper are stored in the public repository Zenodo under DOI [10.5281/zenodo.8033495](10.5281/zenodo.8033495).

## Simulation findings

Figure 2 shows a sample path of $X_t$ over time $t$. The subject dies at age τ = 72.1 (years) with corresponding HRQoL value of $X_{τ-}$ = 0.68. For comparison, a HRQoL value of 0.64 corresponds to a health state similar to home dialysis (rated by patients) [Tor89]. Note that in Figure 2 for $t ≥ τ$ the HRQoL is $X_t = 0$.

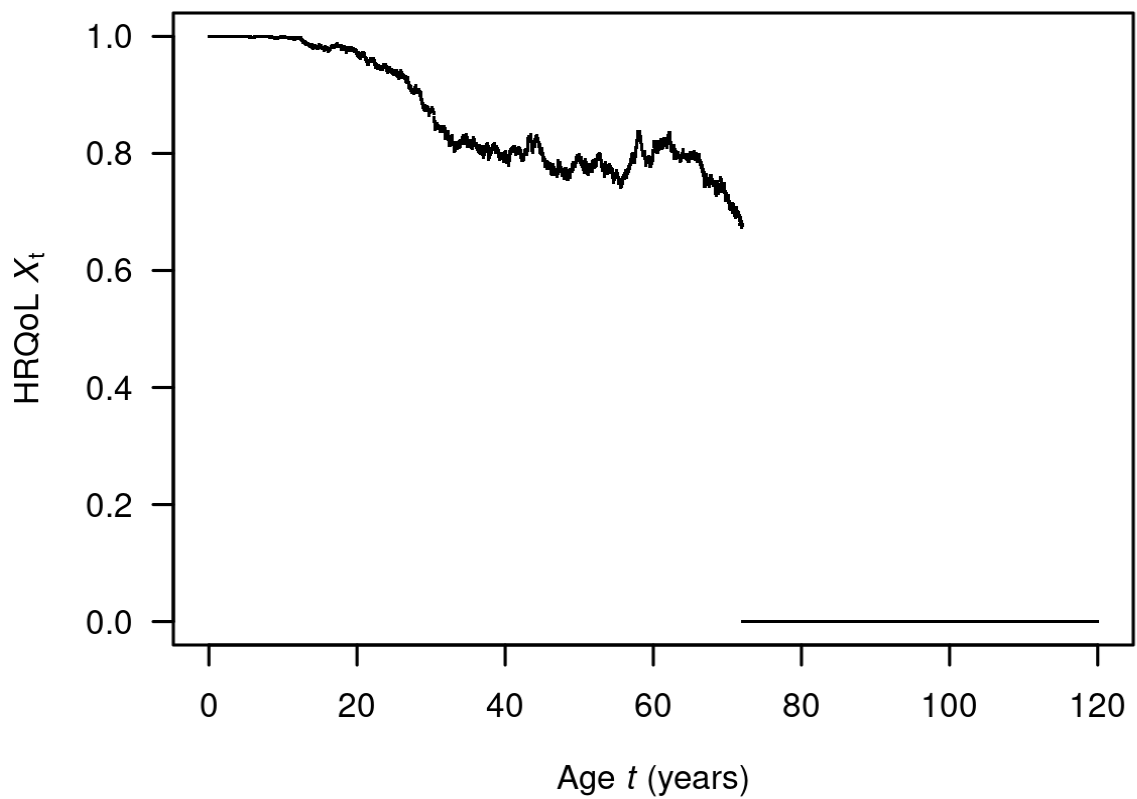

Figure 2: Example path of HRQoL over age for an individual in the simulation.

Median life expectancy in the population of $n$ = 1000 subjects is 83.19 years with interquartile range (IQR) from 74.62 to 88.63 years. For comparison, life expectancy of German women in 2021 was 83.2 years. Median HRQoL at time of death is 0.5797 with IQR from 0.4402 to 0.7050. For comparison, HRQoL values of 0.45, 0.54, 0.70 correspond to health states of being anxious/depressed, having home dialysis (rated by the general public), and moderate angina pectoris, respectively [Tor89].

Figure 3 shows the (pointwise) quartile functions of the $n = 1000$ paths of the stopped process $X^{(\tau)}$.

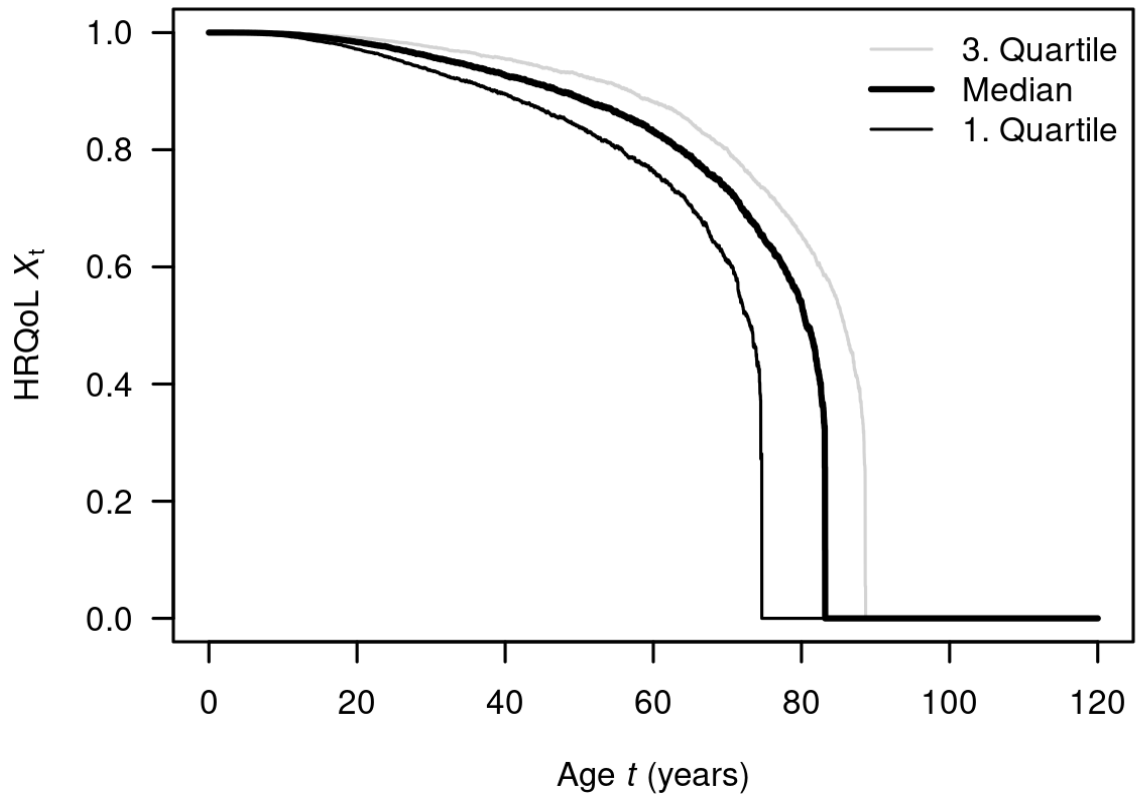

Figure 3: Median and quartiles of the HRQoL in the simulated population of $n = 1000$ people.

Median HALY in the population is 72.23 years with IQR from 66.43 to 76.80 years. For comparison, health life expectancy for German women in 2019 was 72.1 years according to the World Health Organization (for ease of the discussion, healthy life expectancy and HALY are seen as equivalent. For details see [Gol02]).

# Discussion

In this article, we have demonstrated that HRQoL can be modelled as an Ito Stochastic Differential Equation (SDE) with stopping times as age of death. Compared to the German female population, we obtain reasonable epidemiological summary measures with respect to life expectancy and health adjusted life years, which gives a hint for the usefulness of the chosen approach. An advantage of choosing SDEs for modelling population health can be seen in the fact that a very elaborated mathematical theory of SDE including sophisticated numerical treatment is available.

For the time being, we have not elaborated how the parameters of the SDE (1) can be estimated from empirical data. Questions of identifiability of drift and diffusion from given data such as distributions of age-specific HRQoL, life expectancy, and HALE are beyond the scope of this article.

Surveying HRQoL in a group of patients raises a lot of practical questions, for example, which of the many existing instruments for measuring HRQoL should be chosen. The Australian Centre on Quality of Life at Deakin University, Melbourne Campus, maintains a directory of measurement instruments, which comprises several hundred instruments, mostly designed for assessing HRQoL for specific diseases. In this paper we are primarily interested in theoretically modelling HRQoL as a stochastic process that is the solution of an Ito SDE. Hence, practical aspects play a minor role for our considerations here.

Contact:
Ralph Brinks, Chair for Medical Biometry and Epidemiology, Witten/Herdecke University, Faculty of Health/School of Medicine, D-58448 Witten, Germany
ralph.brinks@uni-wh.de